# Unveiling the 3D Morphology of Epitaxial GaAs/AlGaAs Quantum Dots

Yiteng Zhang[1†], Lukas Grünewald[2†], Xin Cao[1], Doaa Abdelbarey[1], Xian Zheng[1], Eddy P. Rugeramigabo[1], Johan Verbeeck[2], Michael Zopf[1,3*], and Fei Ding[1,3]

[1]Institut für Festkörperphysik, Leibniz Universität Hannover, Appelstraße 2, 30167, Hannover, Germany

[2]EMAT, University of Antwerp, Groenenborgerlaan 171, B-2020 Antwerp, Belgium

[3]Laboratorium für Nano- und Quantenengineering, Leibniz Universität Hannover, Schneiderberg 39, 30167, Hannover, Germany

†Authors contributed equally to this work

*Corresponding authors: michael.zopf@fkp.uni-hannover.de

## Abstract

**Strain-free GaAs/AlGaAs semiconductor quantum dots (QDs) grown by droplet etching and nanohole infilling (DENI) are highly promising candidates for the on-demand generation of indistinguishable and entangled photon sources. The spectroscopic fingerprint and quantum optical properties of QDs are significantly influenced by their morphology. The effects of nanohole geometry and infilled material on the exciton binding energies and fine structure splitting are well understood. However, a comprehensive understanding of GaAs/AlGaAs QD morphology remains elusive. To address this, we employ high-resolution scanning transmission electron microscopy (STEM) and reverse engineering through selective chemical etching and atomic force microscopy (AFM). Cross-sectional STEM of uncapped QDs reveals an inverted conical nanohole with Al-rich sidewalls and defect-free interfaces. Subsequent selective chemical etching and AFM measurements further reveal asymmetries in element distribution. This study enhances the understanding of DENI QD morphology and provides a fundamental three-dimensional structural model for simulating and optimizing their optoelectronic properties.**

## Introduction

Semiconductor quantum dots (QDs) embedded within a single-crystalline host matrix represent exceptional sources of non-classical photons, playing a pivotal role in quantum technologies[1-3]. The GaAs/AlGaAs QDs produced through droplet etching and nanohole infilling (DENI) are highly attractive due to their anticipated minimal crystal defects[4] and outstanding optical properties, excelling in key aspects such as pure[5], bright[6,7], and indistinguishable[8,9] single-photon emission with linewidths close to Fourier limit[10-12], as well as strong multi-photon entanglement[13,14]. After enhancing their optical properties through post-growth techniques, such as electrical fields[6,15], strain fields[13,16,17], and optical cavities[18], they can be utilized in a variety of quantum photonic systems, including entanglement swapping[19,20] or quantum key distribution[21]. It is important to note that understanding and controlling the morphology of DENI QDs, in particular their symmetry and material composition, is essential to achieve the desired optical properties. For example, QDs obtained by filling highly symmetric nanoholes show nearly vanishing fine structure splitting and thus improve the fidelity of polarization entanglement[17], while homogeneous QDs produce a narrow wavelength distribution[22]. Intrinsic anisotropic strain fields and in-plane anisotropy of the confinement potential influence the splitting and mixing of heavy and light holes[23]. Smaller QDs tend to weaken the hyperfine interaction caused by the distance between electrons and nuclear spins, and also show a significant inhibitory effect on spin-orbit coupling, thereby extending the spin coherence time[24,25]. In addition, reducing the compositional intermixing of AlGaAs and GaAs in the QDs will further extend the spin coherence time[25,26]. The energy of the confined exciton in DENI QD is mainly determined by the confinement in the growth direction and the infilling amount of GaAs. As a result, different filling amounts realize a wide distribution of exciton emission wavelengths[27-29]. Meanwhile, the weak confinement due to large lateral dimensions, which often exceed the free exciton Bohr radius in GaAs, shortens the exciton radiation lifetime[30]. Therefore, a comprehensive understanding of the three-dimensional (3D) morphology of the DENI-based QD system is essential for optimizing the optical properties intrinsically.

Various techniques are employed to study the morphology of QDs, including *ex-situ* atomic force microscopy (AFM)[28,31], cross-sectional scanning transmission electron microscopy (STEM)[24,32,33], and chemical etching[31,34,35]. AFM provides valuable information about the depth, symmetry, and density of droplet-etched nanoholes in various etching environments or elucidates the facet formation[36] of the Al droplet-etched nanoholes. However, the morphology investigation of filled QDs has received little attention, a critical aspect that directly impacts the comprehension of their optical properties. TEM and STM are effective tools for analyzing crystal strain defects, material composition, and interfacial properties in both strained InGaAs[33] or unstrained droplet epitaxy GaAs[37] QDs respectively. Nonetheless, achieving precise nanoscale localization can be challenging regarding the lower density of DENI nanostructures, thus only a few studies have been conducted on the morphology of DENI QDs[24]. Selective wet etching can complement and thus compensate limitation of AFM and TEM, which can only perform two-dimensional characterization. Etch rate increases around the defect sites, such as dislocations, vacancy agglomerates, or stacking faults, and causes etch pit formation[38,39,40]. In combination with AFM, the 3D alloy composition and distribution of In(Ga)As[31] and SiGe[41] QDs has been successfully obtained, while the composition distributions and conductance distributions of GeSi quantum rings[42] have been detected. This method provides a fast, intuitive, and controllable way to study the QD morphology. It is worth noting that although each method has its limitations, it is promising to achieve a comprehensive understanding of the 3D morphology of GaAs/AlGaAs by using multiple methods that complement each other.

Here, we comprehensively demonstrate the first full-scale 3D morphology of DENI QDs. To do this, we use high-resolution STEM to investigate the cross-section morphology, revealing asymmetric sidewalls and Al-rich regions in the nanohole. The lack of crystal defects and strain suggests that the QDs have coherently crystallized within the nanohole. Selective chemical etching combined with AFM is then employed to further analyze the composition, structure, and distribution in size and symmetry. Our work lays the foundation for morphology analysis of DENI QDs and provides a precise physical model for future studies.

## Results

The GaAs/AlGaAs QDs analyzed in this study are grown using solid-source molecular beam epitaxy with *in-situ* DENI[43] method (supplementary information). The formation of QDs is followed by a drastic mixing of the involved atoms at the interface to the nanohole. This results in element distribution inhomogeneities that affect possible exciton radiative recombination[26]. Therefore, the 'uncapped QDs', which are not contaminated by Al from the capping layer and are easily localized by STEM, are the optimal choice to study the DENI QD morphology using selective chemical etching. To understand the morphology of the nanohole-to-QD structure, we grow a series of uncapped QDs with different filling amounts (Figure 1).

We first obtain nanoholes etched by Al droplets as shown in Figure 1a. The surface of the nanohole is surrounded by an asymmetric (quantum) ring consisting of a larger crescent-shaped part and a separate but higher part[28,36] (cf. solid and dashed arrows in Figure 1a). This asymmetry is attributed to the ripening direction of Al droplet[44]. The openings of nanoholes are subcircular with a slightly shorter width of approximately 68 nm in the [110] orientation compared to 83 nm in the [1-10] orientation. As shown in Figure 1f for the unfilled nanohole (denoted as F0nm), the contour lines in the two orthogonal directions indicate that the as-etched nanoholes are in the shape of an inverted cone, with an average depth of 22 nm in both orientations. The ring shows a rough surface, and the inner wall of the inverted cone exhibits a multidirectional faceted morphology[36]. Additionally, excess Al extends across the substrate surface together with the incident As flux, forming a flat thin layer of AlAs between the nanoholes during the annealing stage.

After the droplet etching process, we introduce different amounts of GaAs material, leading changes in surface morphology from Figure 1b to 1e. The size of the nanohole opening gradually decreases and the volume of the rings increases along the [1-10] orientation. At a filling amount of 0.99 nm (Figure 1d), the nanohole is almost filled, leaving a small shallow pit within the ring. Eventually, when overfilled to 1.99 nm (Figure 1e), the nanohole completely disappears and merges with the ring into an elliptical bump due to the tendency to reduce the surface energy (Figure 1f). The behavior is attributed to the differences in surface area demonstrated by the increased difference in facet angles[36,45] (inset of Figure 1f). The epitaxial growth of GaAs is enhanced in the [1-10] orientation due to preferential migration during the growth process, resulting in an average width of approximately 700 nm along the long axis(Figure 1g). The average height of 3 nm decreases from the center to the periphery, while the outline has smaller pits[28,46] (Figure S 1d). Therefore, a typical GaAs QD is produced, consisting of an elliptical bump capping an asymmetric ring in the upper part and an inverted cone in the lower part.

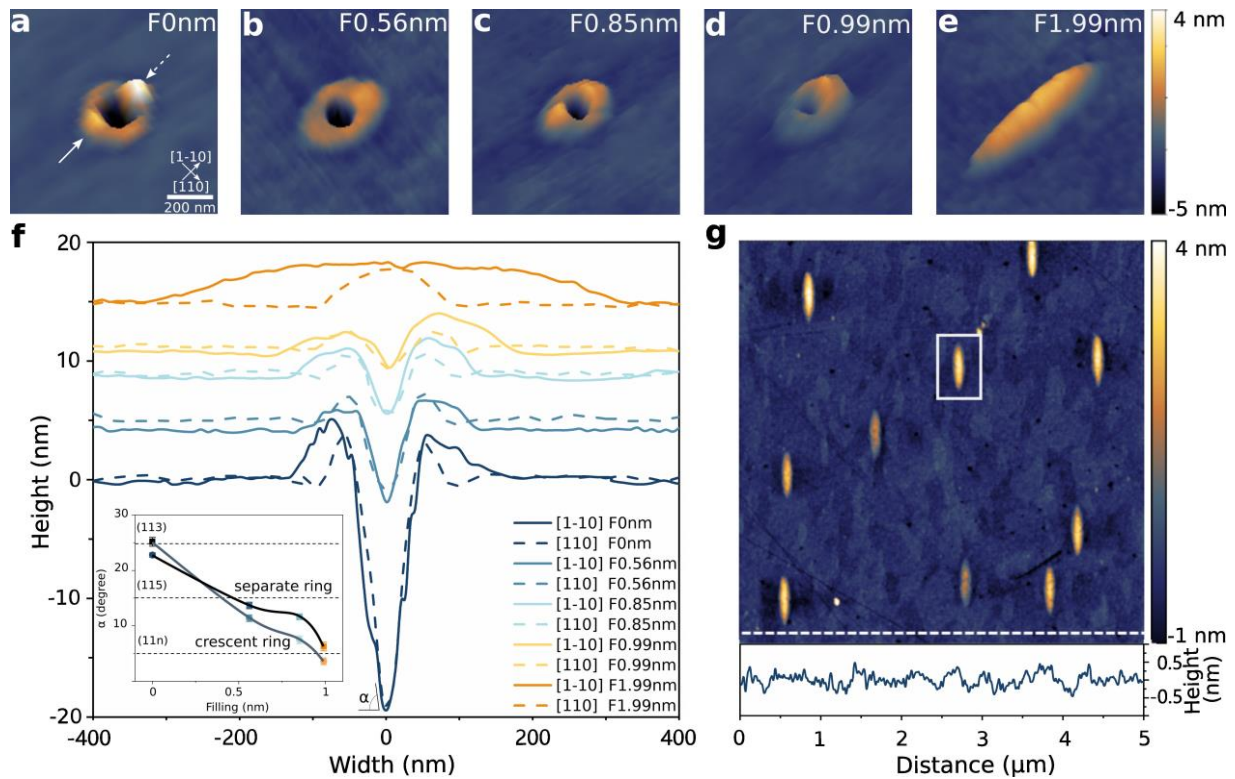

**Figure 1: Morphology of GaAs QDs with varying infilling amounts.** Panels **(a)** to **(e)** illustrate the AFM tilted view (30°) morphology change from nanohole to filled QDs, corresponding to filling amounts of 0 nm, 0.56 nm, 0.85 nm, 0.99 nm, and 1.99 nm, respectively. The nominal amount of GaAs filling is denoted as F$X$nm. The crescent ring faces random orientations, either [1-10] or the opposite. The solid and dashed arrows in **(a)** represent crescent-shaped and separated quantum rings, respectively. **(f)** compares the QD line profile along the [1-10] and [110] orientations. As the amount of infillings increases, the outer diameter of the quantum ring increases from 240 nm to 265 nm, 285 nm, 340 nm, and 700 nm along the [1-10] orientation. The inset illustrates the correlation between the main facet of the nanohole to (001) crystal plane angle α and the amount of filling. A larger facet index (11n) results in a faster growth/etch rate. The F1.99nm sample has a root mean square (RMS) roughness of 0.2 nm, and its surface morphology is shown in **(g)**. The roughness curve along the [110] orientation (dashed white line at the bottom of **(g)**) indicates the layer-by-layer growth of the sample over 5 μm. **(e)** is a magnified view of the QD in a white dotted box in **(g)**.

We randomly select an elliptical bump QD for TEM characterization (supplementary information). The microstructure and chemical composition of the QDs are analyzed by TEM of a cross-section sample as shown in Figure 2. The contour of the QD region is visible in Z-contrast high-angle annular dark-field (HAADF) STEM imaging, where the local image intensity is roughly proportional to the atomic number $Z^{1.7}$ [47,48] (Figure 2a). The QD has an inverted triangular shape with a width of approximately 80 nm at the surface and grows about 25 nm into the Al$_{0.23}$Ga$_{0.77}$As barrier. These dimensions are similar to the nanohole sizes (Figure S 1e), indicating that the cross-section is at the center of the QD. Chemical analysis using energy-dispersive X-ray spectroscopy (EDS) reveals an increased Ga signal and Al depletion in the QD region (Figure 2b), as expected for GaAs relative to $Al_{0.23}Ga_{0.77}As$. The observation mentioned above is also evident in the summed-up EDS spectra (Figure 2c) of the QD and $Al_{0.23}Ga_{0.77}As$ regions. The Ga peak intensities are higher in the triangular QD region (cf. solid arrows in Figure 2c), whereas the Al signal is reduced (dotted arrow in Figure 2c).

An increase in the Al signal is visible on the QD sidewall (Figure 2b), which also explains the reduced HAADF-STEM intensity in Figure 2a (dashed black arrow). The As signal is constant in this field of view. The latter observation may indicate that the sidewall is composed of an AlAs layer, as the nominal As concentration (50 at%) is the same for $Al_{0.23}Ga_{0.77}As$, GaAs, and AlAs (Figure 2b). The effective atomic number $Z_{eff}$[49,50] of AlAs ($Z_{eff}$ = 24.1) is noticeably lower than that of the GaAs QD ($Z_{eff}$ = 32.0) and $Al_{0.23}Ga_{0.77}As$ ($Z_{eff}$ = 30.3), leading to a reduced HAADF-STEM intensity. Additionally, we find that the sidewall thickness of the inverted cone surrounded by AlAs varies, with an Al-rich sidewall region of approximately 5-8 nm and a bottom inverted cone of 15 nm (Figure S 3b). The asymmetric Al signal around the nanohole observed by STEM-EDS (Figure 2b) is in agreement with the asymmetric sidewalls observed by AFM (Figure 1a). The Ga concentration is maximum at the top of the QD 'core' region and gradually decreases to a minimum at the bottom, while the Al concentration exhibits the opposite trend, reaching its highest concentration at the bottom (Figure 2b). These gradients can be explained by the TEM-sample geometry, where the QD core is roughly sliced in the middle along the [1-10] orientation (Figure S 3a). Then, the electron beam detects about half of the [110] orientation nanohole in projection. As a result, the Ga signal decreases toward the bottom of the QD. In contrast, the Al signal may stem (i) from the AlAs sidewall, (ii) the $Al_{0.23}Ga_{0.77}As$ layer, and (iii) possible intermixing of AlAs and GaAs. The latter would result in an $Al_xGa_{1-x}As$ layer around the QD core, but this could not be clarified. A visible O signal is present on the film surface due to sample oxidation. The oxidized layer has a thickness of approximately 5 nm (Figure S 5), and an amorphous structure resulting from natural oxidation of GaAs in air. This structure is similar to the layered structure reported by Toyoshima et al.[51]

Higher-magnification STEM imaging reveals that the QD protrudes approximately 3 nm above the surrounding $Al_{0.23}Ga_{0.77}As$ layer (Figure 2d and 2e), consistent with the statistical results (Figure S 1f). The absence of intensity differences around the QD in the low-angle annular dark-field (LAADF) STEM images (Figure 2d) suggests the absence of strain fields[52,53] at the QD-$Al_{0.23}Ga_{0.77}As$ (or AlAs) interface. The GaAs QD grows coherently inside the nanohole in $Al_{0.23}Ga_{0.77}As$ and is expected to be separated by an AlAs transition layer. In contrast, the cloud-like intensity variations visible in Figure 2d and 2e are caused by slight thickness variations and possible contamination of the TEM sample. The HAADF-STEM image in Figure 2e shows no intensity variations between the $Al_{0.23}Ga_{0.77}As$ layer and QD, which contrasts with Figure 2a. The discrepancy between Figure 2a and 2e can be attributed to the higher electron-beam current used during STEM-EDS mapping and for acquiring the HAADF-STEM image in Figure 2a. The higher signal-to-noise ratio in the latter compared to Figure 2e reveals the slight changes in layer composition resulting in different Z-contrast. High-resolution HAADF-STEM imaging of the QD-$Al_{0.23}Ga_{0.77}As$ interface shows no crystalline defects (Figure 1f). Note that the displayed image was denoised[54]. Together with the LAADF-STEM signal in Figure 2d, which shows no contrast variations indicative of such defects or strain, we conclude that the QD-Al$_{0.23}$Ga$_{0.77}$As interface is highly coherent for the analyzed QD. This defect- and strain-free structure provides a stable environment for subsequent selective etching.

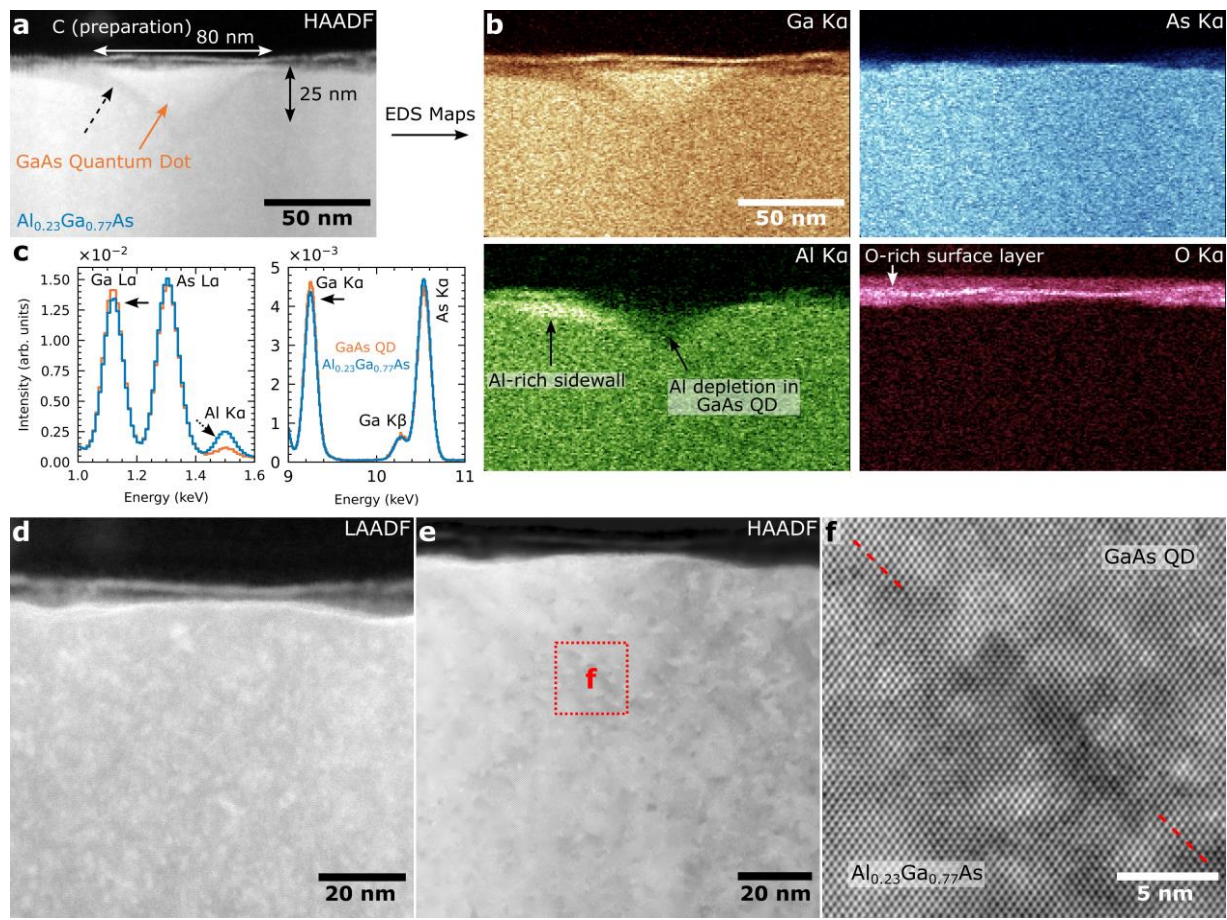

**Figure 2: Microstructural characterization of the QD core by TEM.** (a) Overview cross-section HAADF-STEM image of a 25 nm thin GaAs QD at the surface of an Al$_{0.23}$Ga$_{0.77}$As matrix. The black dashed arrow represents an area of enhanced Al signals, and the yellow arrows represent the QD 'core' region. (b) Elemental maps of selected elements from the region shown in (a). The GaAs QD shows Al depletion, slight Ga enrichment, and no change in As signals. Oxidation of the sample leads to an oxidized surface layer. Al-rich asymmetric sidewalls on the left and right sides can be seen in the Al map. (c) Qualitative comparison of summed-up and normalized EDS spectra from the GaAs QD and Al$_{0.23}$Ga$_{0.77}$As regions. Note the depletion (enrichment) of the Al (Ga) signal in the QD region relative to the matrix, marked with dotted (solid) arrows. The EDS spectra intensities were normalized to the total number of X-ray counts in each spectrum for easier comparison. (d) LAADF-STEM image of the QD showing no visible intensity changes at the QD-matrix interface, indicating no/low strain or other crystalline defects. (e) and (f) Overview and higher-magnification HAADF-STEM image of the QD-matrix interface. The dashed line roughly marks the expected position of the interface. Note that the displayed image is denoised.

To further confirm the 3D morphological features of DENI QDs, we performed a selective chemical etching procedure (supplementary information). Citric acid and hydrogen peroxide can be used for selective etching of GaAs through an oxidation-reduction reaction, known for its high selectivity to GaAs over AlAs[40]. However, the selectivity to AlGaAs can reach up to 100 depending on the Al concentration[35], with slight modifications to etchant ratios causing substantial changes. It is worth noting that the rate of etching is influenced by material defects, crystal orientation, and surface finish[34]. The use of etching solutions can create distinct features, either in etch pits or hillocks[34]. Therefore, the rough surface of elliptical QDs will be more easily etched into pits, which will further enhance etching. TEM analysis revealed Al-rich sidewalls which are expected to prevent etching and exhibit a nanohole-like morphology. The etching morphologies are divided into three stages as shown in Figure 3.

First, Figure 3a shows that the elliptical bump is reduced, revealing distinct, well-defined structures in the middle, a sharp higher bulge (peak A), and a lower bulge (peak B) emerging along the [1-10] orientation. After the second etching, peak A persists and evolves into a crescent-shaped structure surrounding peak B. A new peak (C) appears next to peak B, opposite to peak A (Figure 3b). We believe the surface oxide is rapidly etched away, which enlarges the surface area of the pits, revealing the peaks in sequence. Peaks A and C are distinguishable as taller crescent- and shorter separate-structure. They resemble the quantum ring structure of the nanohole but are opposite in height, indicating that the crescent ring grows faster and merges with the separate ring. This also explains the crossover for the main facet angles in the inset of Figure 1f. At a nanohole fabrication temperature of 635 °C, the presence of facets along the inner wall may result in stacking faults and enhance the etching rate[36,38,39]. The appearance of the hillock peak B further verifies the existence of multi-angle facets on the sidewall near the nanohole opening.

Second, following the third etching cycle, significant erosion occurs in the connection regions where peak B connects to peak A and peak C. This erosion is evidenced by the formation of distinct pits of different shapes surrounding peak B or peak C (Figure 3c). After the fourth etching, the pits merged, resulting in the formation of an annular groove surrounding the central blunt peak B (Figure 3d). In the fifth etching, peak B is etched away (Figure 3e), resulting in a crater-like shape from the top view. The line scans from the E20s to the E40s indicate significant erosion and an increased distance between the center of peak A and peak C (dotted gray line in Figure 3h). The well-defined grooves observed along the sidewall, with an increasing depth correlated to the number of etching cycles, suggest the possible manifestation of asymmetrical composition distribution. Figure 3h E50s shows that the grooves are etched toward the right and correspond to a higher part, indicating that the left side corresponds to the Al-rich sidewall. The Al-rich sidewalls extend to the surface of the separate ring, while peak B represents filled GaAs, based on comparison with the TEM results. The etching rates of the (111)A and (111)B planes in the zinc-blende structure are non-isotropic due to their termination in different atomic layers along the polar stacking direction[55]. However, the main facet at the bottom of the nanohole is dominated by a large angle (111)[36], so the overall etch rate is lower and almost the same in both orientations. The disappearance of peak B is evidence that the bottom sidewall etch rate is lower than that of GaAs.

Third, the surface of the sample becomes rougher after the sixth and seventh etching cycles. Only elliptical platform imprints are present, which change in size from 700 nm (190 nm) to 660 nm (180 nm) in [1-10] ([110]) orientation. The elliptical shape is maintained due to the combination of deposited GaAs and surface AlAs forming $Al_yGa_{1-y}As$ with a high Al content around the quantum ring, and the simultaneous generation of $Al_2O_3$[40] by-products, which prevent etching. The ring-shaped wall surrounding the hole opening is not discernible, as shown in Figures 3f and 3g. It can be inferred from the protruding parts in the middle of the

E60s and E70s curves that the cone tip base has not been completely etched away. Based on the Al signal of TEM and EDS, it can be judged that the inverted cone bottom is an AlAs layer.

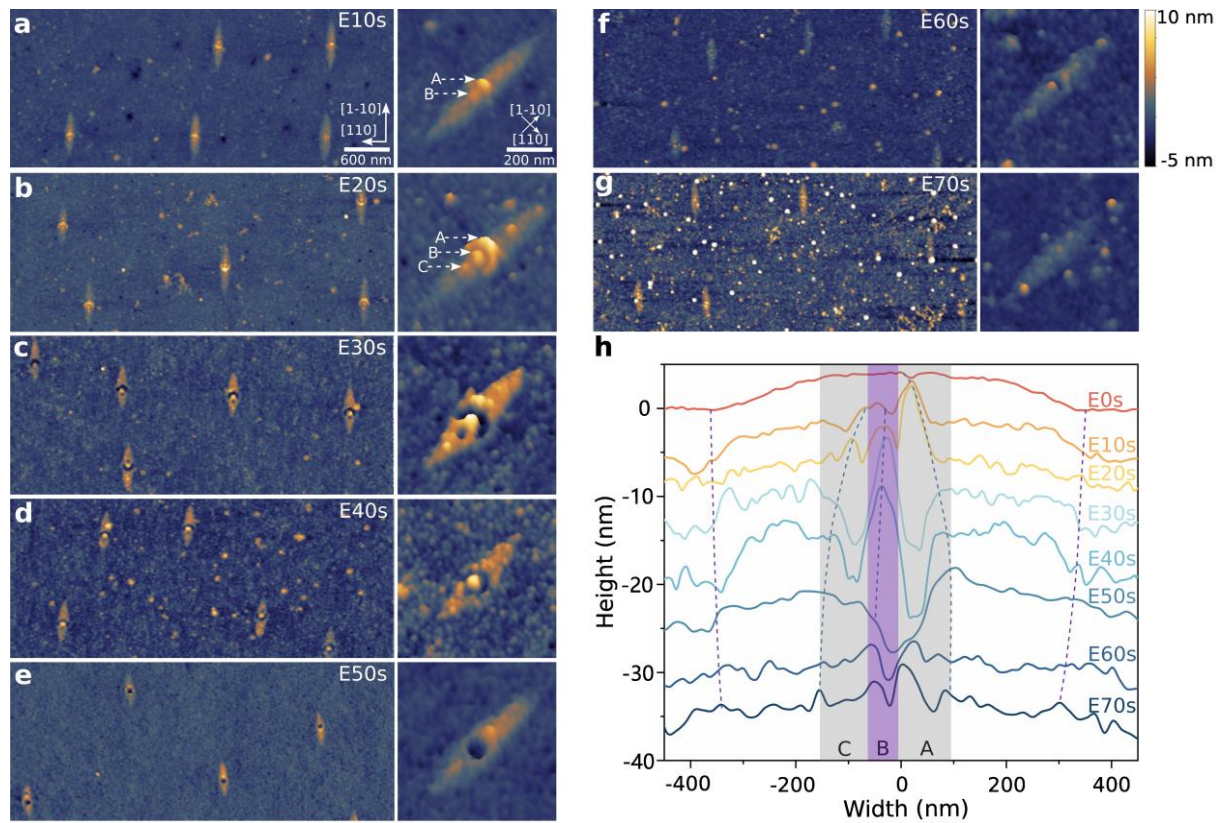

**Figure 3: Morphology evolution of the QD by chemical etching and AFM. (a)** to **(g)** represent the surface morphology changes after etching 10 to 70 seconds (E10s to E70s), respectively. The tilted view (30°) of individual QD images on the right is from the top view QDs **(a)** to **(g)**. **(h)** Line profiles in [1-10] orientation are based on the morphological changes from **(a)** to **(g)**, with the center of the elliptical outline as the benchmark and the height change curve under various etching conditions. The width of the elliptical platform in the [1-10] orientation after etching is shown by the purple dashed line. The gray dashed lines demonstrate the trends of the A, B, and C peaks which are also marked by gray and purple areas, respectively.

## Discussion

Figure 4 shows a true-scale 3D view of the DENI GaAs/AlGaAs QD, its model consistent with all measured data. The etching process indicates that the sidewalls of the majority of nanoholes are not uniformly thick. The thicker portion extends from the bottom of the inverted cone to the separate ring at the surface. Additionally, the etched morphology reveals that the center of the formed elliptical bump is biased toward the crescent ring and does not coincide with the center of the nanohole. The accelerated increase in the lateral dimensions of the grooves suggests that the sidewalls have a characteristic morphology that accelerates etching faster than the GaAs core region, whereas the main large-angle facet inside the nanohole does not increase the etching rate and is therefore attributed to the possible presence of a few nanometer $Al_xGa_{1-x}As$. Thus, it can be concluded that the inverted cone section of the QD is sequentially wrapped by several nanometers of $Al_xGa_{1-x}As$, AlAs of uneven thickness, and $Al_{0.23}Ga_{0.77}As$ barrier layers. The non-uniform thickness of the individual 'shell' layers and the unknown Al content may create free charge noise, which can affect the performance of the photon source.

Experimental evidence suggests that QDs with good optical properties often require high overall symmetry, specifically symmetric quantum rings/nanoholes and filled shallow nanoholes[18,23-26,28].

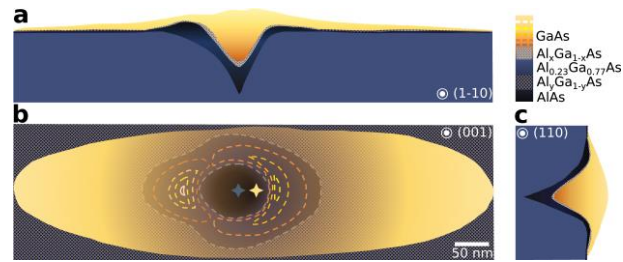

**Figure 4: Three views of GaAs/AlGaAs QDs morphology.** (**a**) and (**c**) are cross-sectional views of the (1-10) and (110) crystal planes respectively, showing the morphology of nanohole and GaAs QDs. For clarity, the [001] orientation in (**a**) and (**c**) is elongated to twice the length of the [110] and [1-10] orientations. (**b**) is a top-perspective view of the QD. The dark blue area in the center represents the nanohole and quantum ring, and the blue star represents the nanohole center. The multiple dashed yellow lines represent the height profile of the quantum ring, while the yellow star represents the center of the elliptical bump. The black regular square represents the high Al content $Al_yGa_{1-y}As$, and the dark blue random square area represents the $Al_xGa_{1-x}As$ layer with lower but unknown Al concentration.

## Conclusion

In conclusion, this study presents the first detailed investigation of the morphology of DENI GaAs/AlGaAs QDs. By combining TEM and AFM of chemical etching, we find that the sidewalls of the inverted conical nanoholes are composed of Al-enriched AlGaAs. The Al content is asymmetrically distributed along the nanoholes and in different crystal directions. Additionally, we investigate the crystallinity of the QD/nanohole interface and reveal a uniform crystalline interface. The QD region contains crystalline and homogeneous GaAs. These observations allow for a detailed physical model of the QD morphology, providing a basis for accurately simulating and deterministically optimizing the electronic optical properties of DENI QDs.

## Acknowledgment

The authors gratefully acknowledge the German Federal Ministry of Education and Research (BMBF) within the projects QR.X (16KISQ015), SemIQON (13N16291), SQuaD (16KISQ117), and QVLS-iLabs: Dip-QT (03ZU1209DD), the European Research Council (MiNet – No. GA101043851), MWK Niedersachsen (QuanTec-76251-1009/2021), and the Deutsche Forschungsgemeinschaft (DFG, German Research Foundation) within the project DI 2013/6-1, as well as under Germany's Excellence Strategy (EXC-2123) Quantum Frontiers (390837967). J.V. and L.G. acknowledge funding from the Flemish government (iBOF-21-085 PERsist). Y. Z. acknowledges the China Scholarship Council (CSC201908370225). We want to thank Stijn van den Broeck for careful TEM sample preparation and Tom Fandrich and Armando Rastelli for fruitful discussions.

## Author contributions

F.D. conceived the project. M.Z. and F.D. supervised the whole research. Y.Z., X. C., and E. R. were responsible for sample growth. L.G. and J.V. were responsible for TEM characterization and data analysis. Y.Z. and X. Z. performed chemical etching and carried out the AFM measurements. Y.Z., L.G., and D.A. wrote the manuscript with input from all other authors.

## Competing financial interests

The authors declare no competing financial interests.

## Data Availability

TEM and AFM data used in this study are available on Zenodo.